# Conspiracy and debunking narratives about COVID-19 origination on Chinese social media: How it started and who is to blame

Version: November 16th, 2020

## Article's Lead

*This paper studies conspiracy and debunking narratives about COVID-19 origination on a major Chinese social media platform, Weibo, from January to April 2020. Popular conspiracies about COVID-19 on Weibo, including that the virus is human-synthesized or a bioweapon, differ substantially from those in the US. They attribute more responsibility to the US than to China, especially following Sino-US confrontations. Compared to conspiracy posts, debunking posts are associated with lower user participation but higher mobilization. Debunking narratives can be more engaging when they come from women and influencers and cite scientists. Our findings suggest that conspiracy narratives can carry highly cultural and political orientations. Correction efforts should consider political motives and identify important stakeholders to reconstruct international dialogues toward intercultural understanding.*

## RESEARCH QUESTIONS

- **Content**. What conspiracy narratives on the origination of COVID-19 are prevalent on Chinese social media over different outbreak phases? How are they similar or different from conspiracy narratives popular in the US? According to these conspiracies, which countries/entities are to blame for the origination of the COVID-19 pandemic?
- **Engagement**. What kind of social media users help propagate conspiracy and debunking posts, and how do they engage with these posts? What debunking strategies are more successful in engaging users?

## ESSAY SUMMARY

- We use the largest to date COVID-19 Weibo corpus to understand the prevalent conspiracy and counteractive narratives regarding COVID-19 origination, over different phases during the pandemic, from January 1 to April 30, 2020.
- Popular conspiracies about COVID-19 on Weibo differ substantially from those in the US. Conspiracies about COVID-19 as human-synthesized or a bioweapon are prevalent on Weibo, especially following Sino-US confrontations.
- Most conspiracy posts on Weibo faulted the US for COVID-19 origination, whereas most debunking posts sought to absolve China from responsibility.
- Debunking conspiracies can be more engaging when they come from women and influencers and cite scientific sources.

## *Argument & Implications*

COVID-19 has garnered a massive amount of conspiracy narratives on social media since January 2020. Conspiracy refers to "an effort to explain some event or practice by reference to the machinations of powerful people, who attempt to conceal their role" (Sunstein & Vermeule, 2009, p.205). In the COVID-19 context, conspiracy centers around virus origination (i.e., who created and spread it). Such misbelief can erode institutional trust, dampen international relations, generate xenophobia, or decrease preventive health behaviors (Swire-Thompson & Lazer, 2020). Conspiracy narratives have been examined in the US (Pew, 2020a, 2020b) and in Europe (Georgiou et al., 2020). For example, prominent narratives promote conspiracy ideation that the US government created the virus, the virus is a Chinese bioweapon (Jamieson & Albarracin, 2020), 5G spreads COVID-19 (Ahmed et al., 2020), or Bill Gates was behind the virus for vaccination programs (Georgiou et al., 2020). So far, no study has examined the evolution of conspiracy narratives in China. Understanding variations of conspiracy narratives across different sociopolitical contexts is imperative in correcting such misinformation and is pivotal in building effective transnational cooperation to mitigate the pandemic.

This study focuses on the Chinese social media context, which, against a backdrop of escalating Sino-US conflicts, has fostered various COVID-19 conspiracies that present a different picture from that of the US and the globe. We examined social media posts that propagate and debunk COVID-19 conspiracies. This paper defines **conspiracy posts** as those that spread conspiracies about the origination of COVID-19. This paper defines **debunking posts** to broadly include any posts that disapprove, disagree and refute such conspiracies, either with or without providing evidence (see Appendix G for examples of conspiracy posts and debunking posts). The debunking posts are classified by their content and not restricted to any particular type of user or source. Overall, this study has three important real-world implications.

The first implication suggests political parties, media, and public agencies to avoid purposefully or inadvertently propagating conspiracy narratives, as they not only misdirect the public's attention during a public health crisis but can also breed long term harm such as declining trust towards governments and authorities (Freeman et al., 2020). As our findings suggest, conspiracy narratives were a direct response to the deteriorating Sino-US relationship, and in turn, debilitated the relationship even further, creating a precarious downward spiral. Conspiracies either covertly or overtly endorsed by the two countries' political figures have exacerbated the problem and devastated international collaborations for global pandemic responses.

     Further, pandemic and conspiracy narratives carry highly contextualized cultural and political assumptions and nuances (Ding & Zhang, 2010; Jovančević & Milićević, 2020). As we show, prominent conspiracies about COVID-19 origination center on either human synthesization or biological weapons on Weibo. By contrast, popular conspiracies concerning 5G, Dr. Fauci, and Bill Gates in the US and elsewhere are seldom mentioned on Weibo. Underlying different conspiratorial arguments are different cultural and political orientations toward technologies and the governments. For example, the Chinese nationalism portraying the US as the political and economic threat is a warrant fueling the bioweapon conspiracy. Correcting such conspiracies thus requires further addressing constructed nationalism. A

practical implication is that efforts on mitigating conspiracy narratives need to work on increasing intercultural and international dialogues to identify common interests and values, and to dispel unfounded claims and misunderstandings. In this regard, we suggest government agencies, media, and educators to work on developing more constructive and unbiased narratives of the pandemic and its global responses.

The second implication informs governmental policy on fighting against the susceptibility to conspiracy beliefs. Just as most conspiracy posts on Weibo faulted the US for COVID-19 origination, most debunking posts sought to absolve China from responsibility. The finding suggests that people may selectively endorse and share debunking messages that support their own group, resulting in an ideologically narrow flow of debunking messages to their followers (Shin & Thorson, 2017). Against the backdrop of increasing Sino-US tension, it is challenging to engage the public when debunking certain conspiracy narratives consistent with one's political or national identity. Communication strategies thus need to facilitate dissolving echo chambers around certain conspiracy narratives that politicize health issues (Del Vicario et al., 2016). For example, inoculation could be an effective strategy to reduce the public's susceptibility to conspiracy beliefs (Roozenbeek & van der Linden, 2019). By giving small doses of conspiracy narratives and explicitly warning the public about how specific political motives (e.g., partisanship, international conflict) fuel each conspiracy narrative, we could help the public become more sophisticated at processing various information on social media.

A final implication of this study concerns platform design about creating effective debunking strategies to counteract conspiracy posts. We showed that users were less engaged (i.e., retweet, like, comment) in debunking posts than conspiracy posts, which echoes previous work that false information diffuses significantly faster, farther, and more broadly than true information on social media (Vosoughi, Roy, & Aral, 2018). Fighting conspiracy is a difficult battle, but our study highlighted that influencers and verified organizational users with a larger following could help draw more user participation to debunking posts. Influencers and organizational users can be considered as critical seeds for disseminating debunking information through online social networks (Rubin, 2017). Social media platforms and public agencies may consider actively enlisting their help in the debunking process.

In sum, we propose the following practical recommendations:
- Political parties, media, and public agencies should avoid citing nationalistic and politically motivated conspiracy narratives and make an effort to dispel conspiracy thinking through increased international dialogues.
- Public communication efforts can consider employing inoculation and media literacy education to decrease susceptibility to conspiratorial thinking.
- Social media platforms need to encourage trusted influencers, organizations, and scientists to disseminate debunking information.

## *Findings*

*Finding 1: Popular conspiracies about the origination of COVID-19 on Chinese social media differ remarkably from those in the US. Conspiracies about COVID-19 as human-synthesized and bioweapon are prevalent on Weibo and these posts attribute more responsibility to the US than to China.*

Figure 1 shows the number of posts that attribute responsibility to the US, China and other entities for each origination type, and responsibility attribution comparing conspiracy posts vs debunking posts. We found that conspiracy origination types that dominate the Chinese social media differ from those in the US and around the globe. In the US or around the globe, conspiracy about 5G, Dr. Fauci, and Bill Gates are prevalent (Pew, 2020a; Goodman & Carmichael, 2020). These conspiracies, however, constitute a small proportion of conspiracies on Weibo (4.95%). Prevalent conspiracies on Weibo focus on whether COVID-19 is deliberately made by country actors in labs or as bioweapons.

Comparing responsibility attribution between debunking posts versus conspiracy posts, we found that people are more likely to debunk conspiracies that blame China while propagating conspiracies that blame the US more frequently ($\chi^2$= 564.29, p<0.01). Responsibility attribution to US and China also substantially differ between conspiracies that talk about the natural/unknown origin of COVID-19 versus those that talk about the deliberative formation of COVID-19. For conspiracy posts that believe that the origin of COVID-19 is natural/unknown, responsibility is attributed more frequently to China (31.07%) compared to origination types that believe COVID-19 is deliberately synthesized by human (15.36%) and used as bioweapons (4.20%).

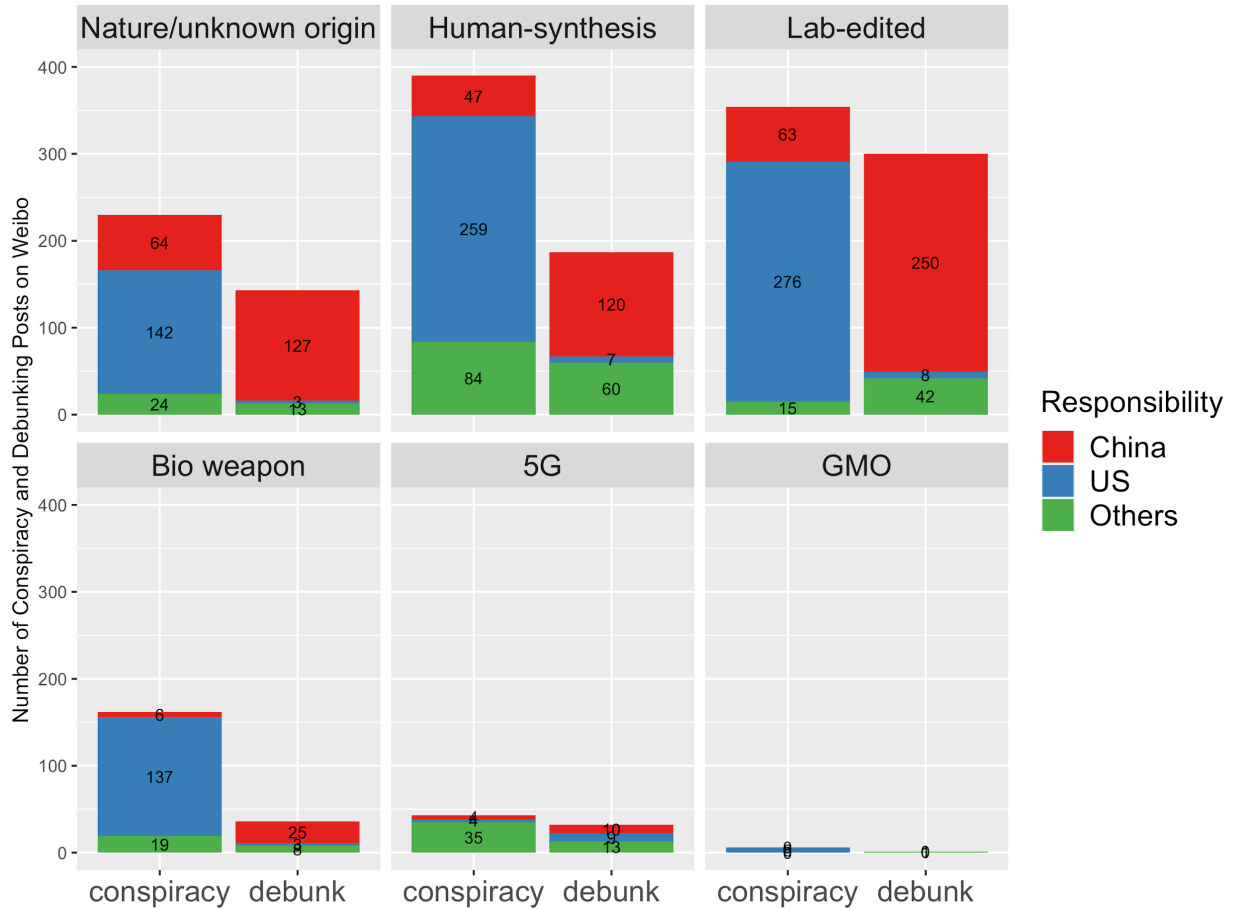

**Figure 1.** *Number of Conspiracy and Debunking Posts of COVID-19 Origination, by Origination Type and Responsibility Attribution*

*Finding 2: Conspiracies that blame the US as the culprit of COVID-19 surged following Sino-US conflicts.*

Conspiracy and debunking narratives as well as responsibility attribution evolved over time with an interesting pattern. Conspiracy posts surged when the US announced sanctioning policies on China such as President Trump tweeted about the China Virus, imposed green card ban, and proposed the 5G cleaning plan on Huawei. While conspiracy posts surged during Sino-US conflicting times, debunking posts surged when China's cases surged around mid-February due to changes in diagnosis testing and when Trump said he would stop using the term China Virus on March 24[th], 2020 (Top Panel). This pattern of how Weibo posts evolve with Sino-US conflicts also persists in terms of responsibility attribution of COVID-19. We found that posts that attribute responsibility to the US for creating COVID-19 virus surged during Sino-US conflicting times (Bottom Panel).

These findings on how conspiracies and responsibility attribution evolved with Sino-US conflicts underscore pandemic as a catalyzer for geopolitical conflicts, nationalism, and misinformation. Our findings echo with the recent literature that stressed that nationalism might harm the equal distribution of COVID-19 vaccines between the Global North and the Global South (Rutschman, 2020). As scholars in psychology explained, the mechanism of "identity-protective cognition" might facilitate the spread of science misperception (Kahan, 2017), as demonstrated by our empirical evidence that conspiracy theories and blaming went hand in hand with Sino-US conflicts. Moreover, the pandemic is reshaping the power structures and international systems between China and U.S., intensifying the Sino-US competition and rivalry (Basu, 2020; Fiona et al., 2020). Escalated narratives focusing on the politics of blaming have grown between China and the U.S. from political speeches to media coverage (Jaworsky and Qiaona, 2020). Science communication has become politicized (Hart, Chinn & Soroka, 2020) and ideological (Wolfe, 2018). As science communication intertwines with political communication (Scheufele, 2014), it is vital to develop mutual understanding and meaningful dialogues between world powers to share responsibility for coping with pandemics and fighting against misinformation.

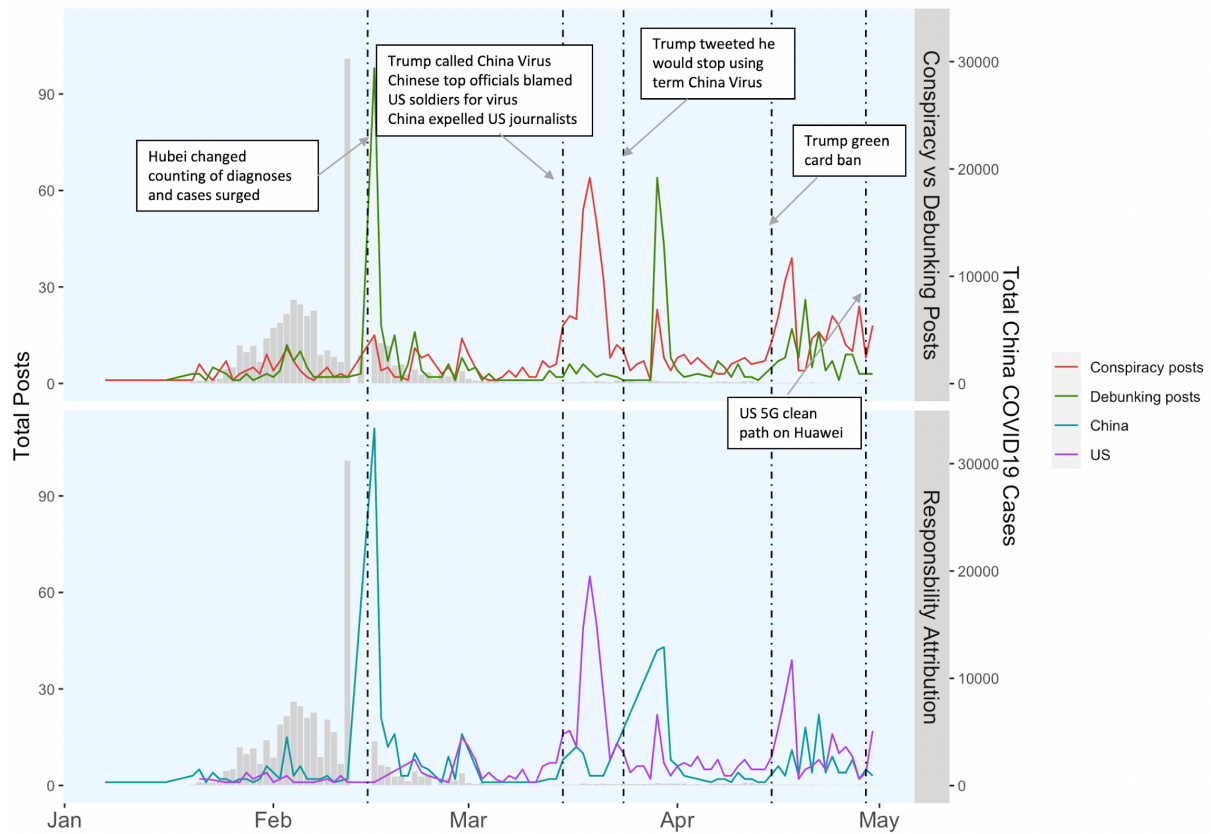

**Figure 2.** *Evolution of Conspiracy and Debunking Narratives, and Responsibility Attribution*

*Finding 3: Men are more likely to propagate and debunk conspiracy posts than women; Influencers and organizational users are overrepresented in the debunking posts than they are in the overall Weibo population.*

We found that the users who propagate conspiracy and who debunk conspiracy are similar in profile. Ordinary users, men, and users with followers between 100-1000 constitute the majority who post conspiracy as well as debunking posts. However, compared with the user profile of the overall Weibo population, a few notable trends emerge.

Among the users propagating conspiracy posts, organizational users are overrepresented (5.19%) than they are in the overall Weibo population (1.46%), while influencers are slightly underrepresented (7.96%) than they are in the overall Weibo population (8.59%). Among the users debunking conspiracy posts, organizational users are again overrepresented (8.51%) than they are in the overall Weibo population, while influencers are overrepresented (10.48%) than they are in the overall Weibo population (8.59%). Men are disproportionately more likely than women to post conspiracy (χ2=108.52, p<0.01) and debunking posts (χ2 =52.57, p<0.01), compared to the gender composition in the overall Weibo population.

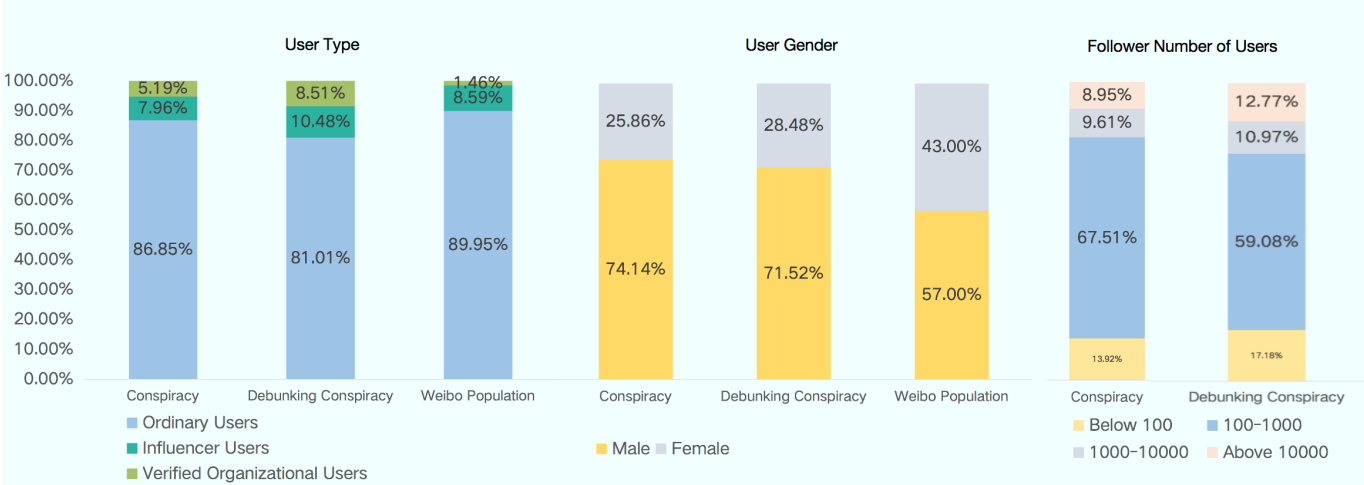

**Figure 3.** *Conspiracy propagation and debunking behaviors by user type[1]*

*Finding 4: Debunking posts have less user engagement than conspiracy posts; However, debunking can be more engaging when it comes from women, influencers and cites scientists.*

In the baseline models with only user attributes and post type (conspiracy vs debunking posts), we found that, compared to conspiracy posts, debunking posts are associated with 10.06% (p=0.07) decrease in user participation (i.e., retweets, likes, comments), but 11.96% percent (p<0.01) increase in user mobilization (i.e., number of @ and hashtags to mobilize others). Although debunking posts are associated with lower participation, we found that the association is moderated by several factors. Within debunking posts, those posted by men received 36.87% less participation than those posted by women [Panel A, right bars], while the same engagement gap for conspiracy posts between men and women was 13.93% [Panel A, left bars]. Within debunking posts, a 10 percent increase in the number of followers is associated with 4.08 percent increase in participation [Panel B, blue line], while for conspiracy posts, a 10 percent increase in the number of followers is associated with 3.05 percent increase in participation [Panel B, red line]. For debunking posts, citing scientists as sources is associated with a higher level of mobilization (20.93%, p=0.02) than those without citing sources [Panel C, right bars]. However, for conspiracy posts, citing scientists is associated with 3.92 percent lower mobilization than those without sources [Panel C, left bars].

---

[1] Note: The data of "the number of followers" for the general Weibo population is not available.

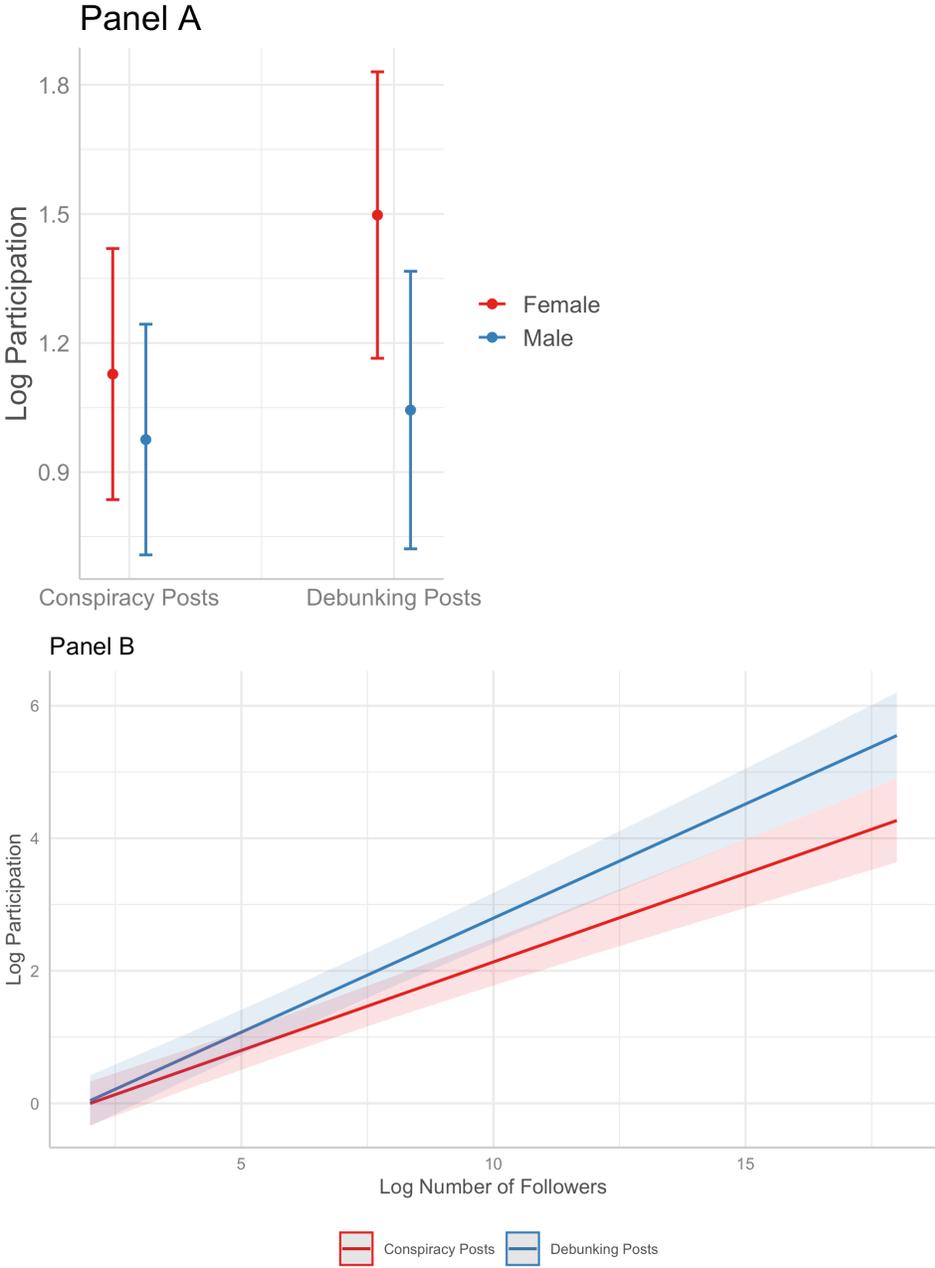

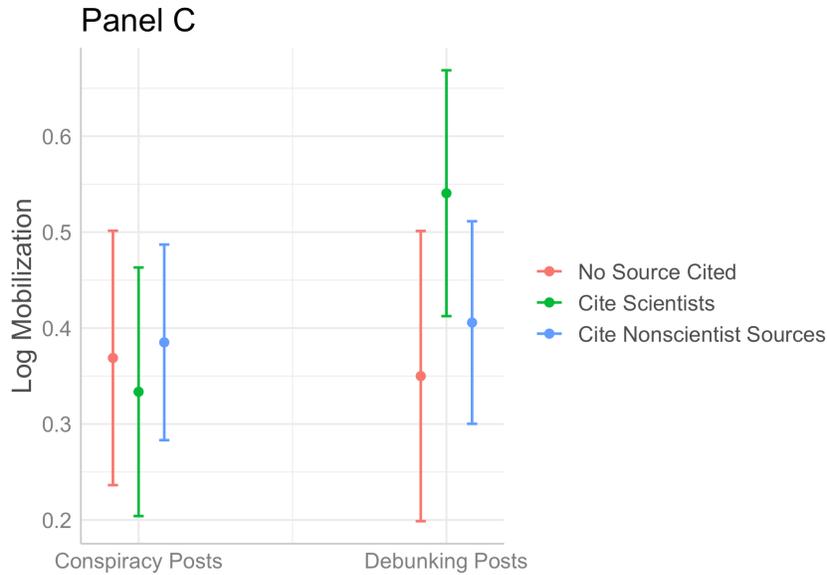

**Figure 4.** *How User Gender (A), Number of Followers (B) and Source Cited (C) Moderate the Associations between Debunking Posts and User Participation and Mobilization[2]*

Our finding that debunking posts by women received more participation than those posted by men responds to a growing body of literature that examines gender differences in public engagement with social media content. For instance, Jia et al. (2018) showed that female online video uploaders were more popular than most male uploaders. The gender differences in how debunking messages were engaged might be due to the language differences women and men use in persuasion (Falk and Mills, 1996). Taking a close reading of the post content by women users that received many reposts, we found that these women used storytelling such as sharing about how COVID-19 has influenced their lives as oversea students. They also used more soft and tentative languages to discuss the COVID-19 origination such as asking for people's mutual understanding about COVID-19 issues, suggesting people to not eat wild animals rather than using hard propaganda languages to attribute responsibility with assertion, which could backfire on audience acceptance about the message senders (Huang, 2018). Through revealing the nuance of these moderators (such as gender), our study provides fruitful future research directions such as investigating how debunking strategies could be matched with specific senders to increase public engagement with science.

---

[2] A 95% confidence interval for the marginal effect of the interaction terms on our dependent variables from the OLS regression are plotted. We took the log form for our dependent variables "participation" and "mobilization" accounting for their skewed distributions. We took the log form for our independent variable the number of followers. For panel B, when log (Number of followers)=0, the intercept of participation for conspiracy posts is -0.53 and the intercept of participation for debunking posts is -0.65. For details on the full regression results, please see Appendix E.

***Methods***

We performed content analyses and regression analyses to examine conspiracy narratives and user engagement. COVID-19 related social media posts were retrieved from Weibo (the Chinese Twitter), with 560 million monthly active users at the end of 2019 (Sina Weibo, 2020). However, Weibo does not provide application programming interface (API) access to the independent researchers, and limits keyword search output to 50 pages (around 1000 posts). To bypass these limitations, we utilized a large Weibo user pool with 250 million users (with bots filtered out) (Shen et al., 2020; Hu, Huang, Chen & Mao, 2020), which accounts for 48.1% of all monthly active Weibo users in 2019 (Sina Weibo, 2020). This user pool was originally built in 2018 and started from a 5 million active Weibo users list collected in our previous studies unrelated to COVID-19 (Li, Cho, Qin & Chen, 2020; Zhang, Nekmat & Chen, 2020), along with a snowball sampling process[3].

---

[3] Using a snowball sampling method, we then retrieved the initial 5 million users' followers and followees (second degree users), the followers and followees of the second-degree users (third degree users) and so on until no new users appear. This snowball process resulted in a pool of 250 million users (with bots filtered out) (Shen et al., 2020).

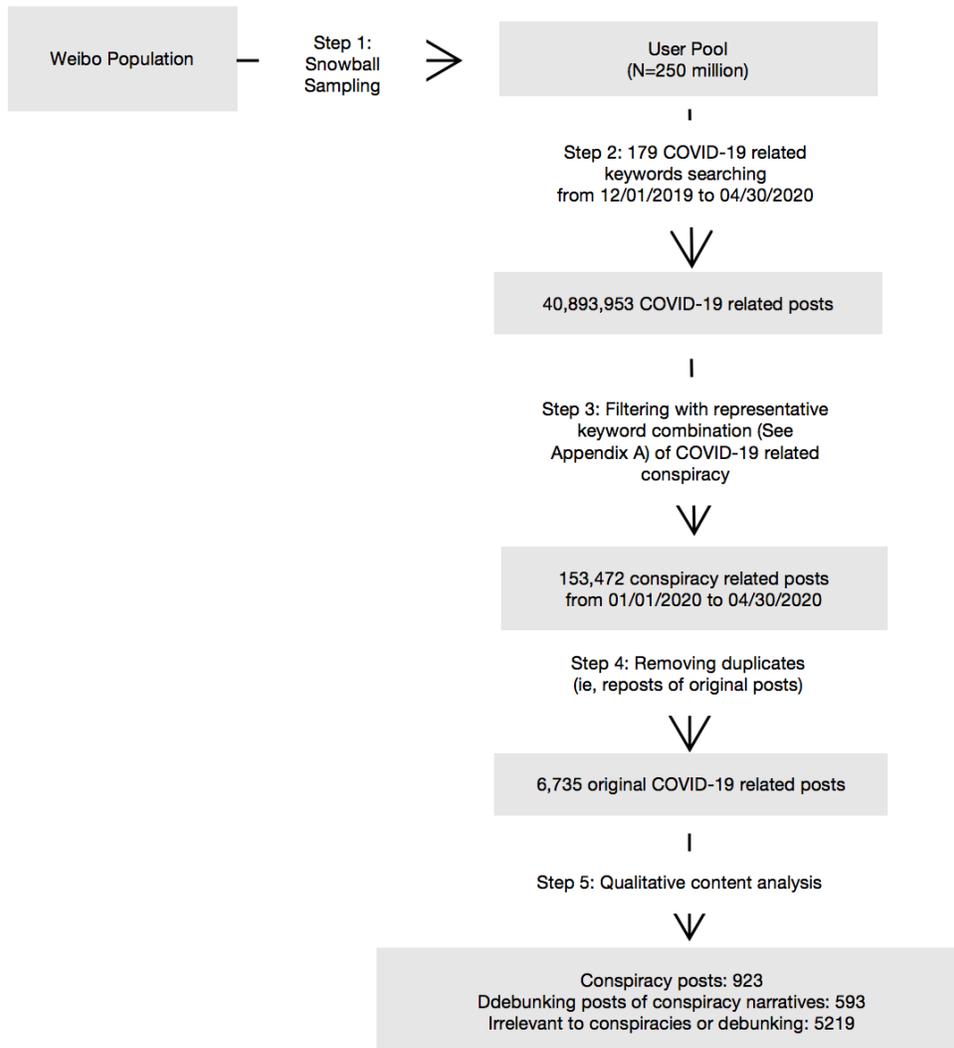

Figure 5. *Weibo data collection and content analysis procedure*

From the user pool, we retrieved COVID-19 related posts using a comprehensive list of 179 keywords (for a complete list, see Hu, Huang, Chen & Mao, 2020). After removing duplicates, we obtained a main corpus of 40,893,953 COVID-19 related Weibo posts between December 1, 2019 (the date of the first known COVID-19 case) and April 30, 2020. Drawing from academic, government and news resources[4], we found 35 COVID-19 conspiracy theories (e.g., "5G spreads virus", "China utilizes COVID-19 to paralyze the Western economy"). We summarized keyword combinations for each specific conspiracy narrative (see Appendix A) via close observation of their relevant posts on Weibo, along with several rounds of back-and-forth discussions. These keyword combinations were adopted to filter the COVID-19 corpus, yielding 153,472 posts from Jan 1, 2020 to April 30, 2020. We removed duplicate posts and reposts as practiced by other studies (Shen et al., 2020; González-Ibánez et al., 2011), because (1) we focus on the narrative of the person who initiated the conspiracy and (2) we study the number of retweets of a post as a dependent variable. The final dataset contained 6,735 unique original Weibo posts (original posts are posts that start threads, not reposts or comments) about COVID-19 conspiracies dated from Jan 1, 2020 to April 30, 2020. These 6735 original posts reached a large audience, generating 31,421 reposts, 260,355 likes, and 38,075 comments.

---
[4] We drew from 1) earlier research on COVID-19 conspiracies (Leng et al., 2020; Imhoff & Lamberty, 2020), 2) Chinese fact-checking websites (e.g., Tencent Jiaozhen), 3) news website (e.g., BBC), and 4) government websites (e.g., Embassy of People's Republic of China in Germany).

We developed a comprehensive coding scheme to manually annotate each post based on four dimensions: post types, origination types, responsibility attribution, and sources cited (See Appendix B for coding scheme). **Post types** focus on distinguishing posts that propagate conspiracies vs disapprove/refute conspiracies. **Origination types** concern the various theories on the origination of COVID-19 such as whether the source of COVID-19 is unknown or made by human actors. **Responsibility attribution** concerns the country or entities who are pointed out in a post as responsible for causing the COVID-19 pandemic. **Sources cited** concerns the type of sources cited in a post such as from government, scientists, media and so on. Six Chinese native speakers were trained and coded the posts independently and satisfied inter-coder reliabilities were achieved (see Appendix C).

To examine the relationship between post type, origination types, responsibility attribution, and sources cited, we first calculated the frequency of each category (See Appendix C). To examine the association between post type and user engagement, two regression models were conducted. In the models, our two dependent variables are post participation including like number, repost number and comment number and post mobilization including number of @ and hashtags in a post. In our baseline model, our independent variables include post type (conspiracy or debunking), user gender, user type, geolocation (Hubei or outside Hubei), emotional factors (i.e., emotion score, emotion polarity, and emotion types) were calculated based on previous practices (Zhang et al., 2017; Zhao et al., 2016), and length of a post (See Appendix D). In the full model, we added hand-coded variables such as origination types, responsibility attribution, source cited in addition to baseline variables. To examine what debunking strategies might be associated with variation in user engagement, our full model also included the interaction of debunking and source cited, debunking and origination types, debunking and user attributes (See Appendix E).

We discussed the limitations of our study in Appendix H about whether censorship could influence our findings, disclaimer on causality, and generalizability of our findings to other media systems.

*Bibliography*
Ahmed, W., Vidal-Alaball, J., Downing, J., & Seguí, F. L. (2020). COVID-19 and the 5G conspiracy theory: Social network analysis of Twitter data. *Journal of Medical Internet Research*, 22(5), e19458. https://doi.org/10.2196/19458

Basu, T. (2020). Sino-US Disorder: Pandemic, power and policy perspectives in Indo-Pacific. *Journal of Asian Economic Integration*, 2(2), 159–179. https://doi.org/10.1177/2631684620940448

Fiona, H., Tanvi, M., Amanda, S., Mireya, S., & Constanze, S. (2020, July 1). Balancing act: Major powers and the global responses to US-China great power competition. *FOREIGN POLICY at BROOKINGS*. https://www.brookings.edu/research/balancing-act-major-powers-and-the-global-response-to-us-china-great-power-competition.pdf

Carli, L. L. (1990). Gender, language, and influence. *Journal of Personality and Social Psychology*, 59(5), 941–951. https://doi.org/10.1037/0022-3514.59.5.941

Del Vicario, M., Bessi, A., Zollo, F., Petroni, F., Scala, A., Caldarelli, G., ... & Quattrociocchi, W. (2016). The spreading of misinformation online. *Proceedings of the National Academy of Sciences*, 113(3), 554–559. https://doi.org/10.1073/pnas.1517441113

Ding, H., & Zhang, J. (2010). Social media and participatory risk communication during the H1N1 flu epidemic: A comparative study of the United States and China. *China Media Research*, 6(4), 80-91. https://d1wqtxts1xzle7.cloudfront.net/6229934/ChinaMediaResearch.pdf

Falk, E., & Mills, J. (1996). Why sexist language affects persuasion: The role of homophily, intended audience, and offense. *Women and Language*, 19(2), 36-44. https://www.questia.com/library/journal/1G1-19265426

Freeman, D., Waite, F., Rosebrock, L., Petit, A., Causier, C., East, A., ... & Bold, E. (2020). Coronavirus conspiracy beliefs, mistrust, and compliance with government guidelines in England. *Psychological Medicine*, 1-13. https://doi.org/10.1017/S0033291720001890

Fu, K. W., & Zhu, Y. (2020). Did the world overlook the media's early warning of COVID-19?. *Journal of Risk Research*, 1-5. https://doi.org/10.1080/13669877.2020.1756380

Georgiou, N., Delfabbro, P., & Balzan, R. (2020). COVID-19-related conspiracy beliefs and their relationship with perceived stress and pre-existing conspiracy beliefs. *Personality and Individual Differences*, 166, 110201. https://doi.org/10.1016/j.paid.2020.110201

González-Ibánez, R., Muresan, S., & Wacholder, N. (2011, June 19). Identifying sarcasm in Twitter: a closer look. In *Proceedings of the 49th Annual Meeting of the Association for Computational Linguistics: Human Language Technologies* (pp. 581-586). https://dl.acm.org/doi/10.5555/2002736.2002850

Goodman, J. & Carmichael, F. (2020, June 27). Coronavirus: 5G and microchip conspiracies around the world. *BBC News*. https://www.bbc.com/news/53191523

Hart, P. Sol, Sedona Chinn, and Stuart Soroka. 2020. Politicization and polarization in Covid-19 news coverage. *Science Communication,* 42(5), 679–697. https://doi.org/10.1177/1075547020950735

Huang, H. (2018). The pathology of hard propaganda. *The Journal of Politics*, 80(3), 1034-1038. https://dl.acm.org/doi/abs/10.1086/696863

Hu, Y., Huang, H., Chen, A., & Mao, X. L. (2020). Weibo-COV: A Large-Scale COVID-19 Tweets Dataset from Webo. https://arXiv:2005.09174

Imhoff, R., & Lamberty, P. (2020). A bioweapon or a hoax? The link between distinct conspiracy beliefs about the Coronavirus disease (COVID-19) outbreak and pandemic behavior. *Social Psychological and Personality Science*, 11(8), 1110-1118. https://doi.org/10.1177/1948550620934692

Jamieson, K. H., & Albarracin, D. (2020). The relation between media consumption and misinformation at the outset of the SARS-CoV-2 Pandemic in the US. *The Harvard Kennedy School Misinformation Review*, 1(2), 1-22. https://206.191.184.172/handle/1/42661740

Jaworsky, B. N., & Qiaoan, R. (2020). The politics of blaming: The narrative battle between China and the US over COVID-19. *Journal of Chinese Political Science*, 1-21. https://doi.org/10.1007/s11366-020-09690-8

Jovančević, A., & Milićević, N. (2020). Optimism-pessimism, conspiracy theories and general trust as factors contributing to COVID-19 related behavior–A cross-cultural study. *Personality and Individual Differences*, 167, 110216. https://doi.org/10.1016/j.paid.2020.110216

Jia, A. L., Shen, S., Li, D., & Chen, S. (2018). Predicting the implicit and the explicit video popularity in a user generated content site with enhanced social features. *Computer Networks*, 140, 112-125. https://doi.org/10.1016/j.comnet.2018.05.004

Kahan, Dan M. (2017,October 2). Misinformation and Identity-Protective Cognition. *Yale Law & Economics Research Paper*. https://ssrn.com/abstract=3046603

Leng, Y., Zhai, Y., Sun, S., Wu, Y., Selzer, J., Strover, S., ... & Ding, Y. (2020). Misinformation during the COVID-19 outbreak in China: Cultural, social and political entanglements. *IEEE TRANSACTION ON BIG DATA*, Manuscript submitted for publication.

Li, P., Cho, H., Qin, Y., & Chen, A. (2020). #MeToo as a connective movement: Examining the frames adopted in the anti-sexual harassment movement in China. *Social Science Computer Review*, 0894439320956790. https://doi.org/10.1177/0894439320956790

Pew Research Report. (2020a, June 29). Three Months In, Many Americans See Exaggeration, Conspiracy Theories and Partisanship in COVID-19 News. *Pew Research Center.* https://www.journalism.org/2020/06/29/three-months-in-many-americans-see-exaggeration-conspiracy-theories-and-partisanship-in-covid-19-news/

Pew Research Report. (2020b, July 30). Americans Fault China for Its Role in the Spread of COVID-19. *Pew Research Center.* https://www.pewresearch.org/global/2020/07/30/americans-fault-china-for-its-role-in-the-spread-of-covid-19/

Roozenbeek, J., & Van Der Linden, S. (2019). The fake news game: Actively inoculating against the risk of misinformation. *Journal of Risk Research*, 22(5), 570-580. https://doi.org/10.1080/13669877.2018.1443491

Rubin, V. L. (2017). Deception detection and rumor debunking for social media. In *The SAGE Handbook of Social Media Research Methods* (p. 342). London: SAGE.

Santos Rutschman, A. (2020). The COVID-19 vaccine race: Intellectual property, collaboration(s), nationalism and misinformation. *Washington University Journal of Law and Policy*, 64. https://ssrn.com/abstract=3656929

Scheufele, D. A. (2014). Science communication as political communication. *Proceedings of the National Academy of Sciences*, 111(Supplement 4), 13585-13592. https://doi.org/10.1073/pnas.1317516111

Shen, C., Chen, A., Luo, C., Zhang, J., Feng, B., & Liao, W. (2020). Using reports of symptoms and diagnoses on social media to predict COVID-19 case counts in mainland China: Observational infoveillance study. *Journal of Medical Internet Research,* 22(5), e19421. https://doi.org/10.2196/19421

Shin, J., & Thorson, K. (2017). Partisan selective sharing: The biased diffusion of fact-checking messages on social media. *Journal of Communication,* 67(2), 233-255. https://doi.org/10.1111/jcom.12284

Sina Weibo. (2020, April 19). Weibo Reports First Quarter 2020 Unaudited Financial Results. *Sina Corporation*. http://ir.weibo.com/node/7726/html

Sunstein, C. R., & Vermeule, A. (2009). Conspiracy theories: Causes and cures. *Journal of Political Philosophy*, 17(2), 202-227. https://doi.org/10.1111/j.1467-9760.2008.00325.x

Swire-Thompson, B., & Lazer, D. (2020). Public health and online misinformation: Challenges and recommendations. *Annual Review of Public Health*, 41, 433-451. https://doi.org/abs/10.1146/annurev-publhealth-040119-094127

Vosoughi, S., Roy, D., & Aral, S. (2018). The spread of true and false news online. *Science,* 359(6380), 1146–1151. https://doi.org/10.1126/science.aap9559

Wolfe, Audra J. (2018). *Freedom's Laboratory: The Cold War Struggle for the Soul of Science*. Baltimore: Johns Hopkins University Press

Zhang, L., Xu, L., & Zhang, W. (2017). Social media as amplification station: Factors that influence the speed of online public response to health emergencies. *Asian Journal of Communication*, 27(3), 322-338. https://doi.org/10.1080/01292986.2017.1290124

Zhang, X., Nekmat, E., & Chen, A. (2020). Crisis collective memory making on social media: A case study of three Chinese crises on Weibo. *Public Relations Review*, 46(4), 101960. https://doi.org/10.1016/j.pubrev.2020.101960

Zhao, N., Jiao, D., Bai, S., & Zhu, T. (2016). Evaluating the validity of simplified Chinese version of LIWC in detecting psychological expressions in short texts on social network services. *PLoS One*, 11(6), e0157947. https://doi.org/10.1371/journal.pone.0157947

***Appendix***

*A. COVID-19 conspiracy theories and their representative keyword combination*

| Conspiracy Narratives | Representative Keywords | Translation |
| --- | --- | --- |
| 震惊哈佛大学教授：新型冠状病毒诞生于人为基因改造 | 新冠 AND人为/新冠AND基因改造 | COVID-19 AND Human Synthesized/COVID-19 AND Gene Editing |
| 武汉病毒所女研究生黄燕玲是新冠肺炎零号病人 | 武汉病毒所AND零号病人/黄燕玲AND零号病人/女研究生 AND零号病人 | Wuhan Institute of Virology AND Index Case/Huang Yanling AND Index Case/Female postgraduate AND Index Case |
| 新型冠状病毒是由实验室制造的生物武器 | 新型冠状AND实验室AND武器 | COVID-19 AND Lab AND Weapon |
| 俄美国疾控中心确认新冠病毒源头是美国 | 病毒源头AND美国 | Virus Source AND USA |
| 台湾专家：根据一篇论文可以得出新冠病毒源头是美国 | 病毒源头AND美国 | Virus Source AND USA |
| 俄罗斯科学家验证新冠病毒为人工合成病毒 | 新冠病毒AND人工合成 | COVID-19 AND Human Synthesized |
| 美国的电子烟肺炎是新型冠状病毒导致 | 电子烟肺炎AND新型冠状 | E-cigarette pneumonia AND COVID-19 |
| 美国疾控中心确认新冠病毒源头是美国 | 病毒源头AND美国 | Virus Source AND USA |
| 华裔教授因新冠研究被灭口？ | 华裔教授AND新冠研究AND灭口 | Chinese professor AND killed AND research on COVID-19 |
| 中国的新型冠状疫情是美国发动的秘密武器，有助于美国制造业复苏……美国病毒战的目的，不但是贸易战，更是毁我长城灭我中华 | 新型冠状AND秘密武器/新型冠状AND病毒战 | COVID-19 AND secret weapon/COVID-19 AND war |
| 2019新型冠状病毒棘突蛋白中含有独特的插入序列，并与艾滋病毒的HIV-1 gp120和Gag蛋白有奇特的相似性 | 新型冠状病毒 AND gp120 or Gag | COVID-19 AND gp120/COVID-19 AND Gap |
| 中国早在2019年11月中旬就获悉疫情爆发，将有关信息隐瞒45天 | 中国AND信息AND隐瞒 | China AND COVID-19 AND Hide |
| 中国长时间隐瞒新冠疫情爆发真相，才导致全球疫情爆发 | 中国AND隐瞒AND疫情AND爆发 | China AND hide AND Epidemic AND Outbreak |
| 台湾早在2019年12月31日就向世卫组织发出关于新冠肺炎人传人的警告，但未获得重视 | 台湾AND人传人AND警告 | Taiwan AND Person-to-person AND Warning |
| 新型冠状病毒早在2018年就被发现 | 新型冠状病毒AND 2018AND发现/新冠状AND 2018/新肺炎AND2018 | COVID-19 AND 2018 AND Discover/COVID-19 2018 |
| 中国长时间隐瞒新冠疫情爆发真相，才导致全球疫情爆发 | 隐瞒AND疫情AND爆发 | Hide AND Epidemic AND Outbreak |
| 中国为了隐瞒疫情爆发，逮捕了最早向世人示警的医生 | 隐瞒疫情AND逮捕AND医生 | Epidemic Hide AND Arrest AND Doctor |
| 中国隐瞒并美化了新冠肺炎确诊和死亡人数 | 中国AND隐瞒AND确诊人数or 死亡人数 | Hide AND Confirmed Cases OR death Cases |
| 中国操纵世界卫生组织，以确保其不会批评中国 | 中国AND 操纵AND世界卫生组织 | China AND Manipulate AND WHO |
| 中国阻止台湾加入世卫组织，危害台湾人的健康 | 中国AND阻止AND台湾AND 加入世卫组织 | China AND Prevent AND Taiwan AND join WHO |
| 中国帮助其他国家抗疫只是为了扩大地缘政治影响力 | 中国AND抗疫AND扩大地缘政治影响力 | China AND Fights the epidemic AND Expands geopolitical influence |
| 中国利用新冠病毒使西方经济瘫痪 | 中国AND新冠病毒AND西方经济瘫痪 | China AND COVID-19 AND Paralyze Western Economy |
| 华大基因出卖中国人的基因信息，美国人针对中国人的基因投放病毒 | 新冠状AND基因武器OR病毒是美国人工合成的OR"可精准攻击华人 | COVID-19 AND Gene Weapon/ Human synthesized Virus AND USA/Accurately attack Chinese |
| 意大利病毒；中国疫情暴发前病毒或已在意大利传播了 | 意大利病毒/ 中国疫情暴发前病毒或已在意大利传播了 | Italy Virus/Virus spread in Italy before outbreak in China |
| 5G传播病毒 | 5G AND 传播病毒 | 5G Spread Virus |
| 拉脱维亚发明冠状病毒 | 拉脱维亚AND发明病毒 | Latvia invented the virus |
| 病毒研究专家石正丽所在的P4實驗室管理不善，涉嫌为泄漏新型冠状病毒的源头； 武汉病毒研究所研究员石正丽曾参与美国科研机构主导的人造冠状病毒研究 | 石正丽 AND人造 AND病毒 | Shi Zhengli AND Human synthesized AND Virus |
| 2020年1月31日，美国共和党参议员汤姆·科顿，认为病毒是武汉实验室泄露的生化武器 | 汤姆·科顿 AND 武汉病毒实验室 AND 生化武器 | Tom Cotton AND Biological weapon AND Wuhan Institute of Virology |
| 加拿大P4实验室国家微生物实验室 | 加拿大 AND P4实验室 AND 病毒 | Canada AND P4 Lab AND Virus |
| 新冠病毒是中国间谍从温尼伯的國家微生物實驗室偷走 | 新冠病毒AND温尼伯AND偷走 | COVID-19 AND Winnipeg AND Stolen |
| 比尔·盖茨被污蔑为"新冠病毒制造者"：为从疫苗中牟利； 比尔·盖茨"试图利用疫情夺取全球的卫生系统； 比尔盖茨通过病毒谋杀人类控制人口；比尔盖茨在疫苗中植入微芯片监控人类 | 比尔盖茨 AND 新冠病毒制造者/比尔盖茨AND疫苗牟利/ 比尔·盖茨AND试图利用疫情夺取全球的卫生系统/比尔盖茨AND通过病毒谋杀人类控制人口 | Bill Gates AND COVID-19 inventor/Bill Gatez AND Vaccine AND Profit/Bill Gates AND epidemic AND take the global health system/Bill Gates AND viruses AND murder OR control population |
| 罗曼诺夫，新型冠状病毒可能起源于美国， 马里兰州迪特里克堡的美国军细菌实验室 | 新型冠状病毒AND马里兰州迪特里克堡/新型冠状病毒AND美国军细菌实验室 | COVID-19 AND Fort Dietrich, Maryland/COVID-19 AND U.S. Army Bacteria Laboratory |
| 新冠病毒是由中国的实验室培养而来 | 新冠病毒AND中国AND实验室培养 | COVID-19 AND Chinese Laboratory AND Nurture |
| 新冠病毒源自中国武汉病毒研究所实验室事故 | 新冠病毒AND中国武汉病毒研究所实验室AND事故 | COVID-19 AND Lab AND Accident AND Wuhan Institute of Virology |
| 中国利用新冠病毒使西方经济瘫痪 | 中国利用新冠病毒使西方经济瘫痪 | China AND Utilize AND COVID-19 AND paralyze the Western economy |

*B. Coding scheme with definitions and operationalizations*

The table below introduced the definition and attributes for our four content analysis variables: post types, origination types, responsibility attribution, and sources cited. **Post types** include: 1) conspiracy posts, 2) debunking posts that correct conspiracy narratives, and 3) irrelevant to conspiracies or debunking: others posts which contain conspiracy keywords but are not specifically about COVID-19 origination (e.g. conspiracy or debunking posts about local government corruption). **Origination types** concern the various theories on the origination of COVID-19, including: (1) COVID-19 came from nature or some unknown origin (nature/unknown origin); (2) COVID-19 was entirely developed by humans (human synthesis), (3) COVID-19 was the result of modifying the genes of one or more natural organisms (lab-edited), (4) COVID-19 was developed as a biological weapon (bioweapon), (5) COVID-19 is caused by 5G (5G), (6) Smoking e-cigarettes caused COVID-19 (e-cigarette), (7) Genetically modified crops caused COVID-19 (GMO), and (8) Others. **Responsibility attribution** concerns the country or entities who are responsible for causing the COVID-19 pandemic as pointed out in a post, including: (1) China, (2) The United States, (3) Japan, (4) Serbia, (5) other European countries, (6) Bill Gates, (7) other countries (outside of the above-mentioned countries), and (8) no clear responsibility attribution. **Sources cited** include: (1) government sources (documents, officials, organizations) (2) scientific scholars, (3) celebrities, (4) ordinary people, (5) foreign media, (6) Chinese media, (7) industry/companies, (8) non-governmental organizations, (9) others (outside of the above categories), and (10) no sources. For origination type, responsibility attribution and sources cited, coders were allowed to select multiple categories simultaneously.

| Category | Definition | Operationalization |
| --- | --- | --- |
| Post types | ・Whether a post is relevant to conspiracy or not; if so, whether it debunks conspiracy or is about conspiracy | ・Conspiracy posts<br>・Debunking posts of conspiracy narratives<br>・Irrelevant to conspiracies or debunking |
| Origination types | ・Different conspiracies regarding the origination of coronavirus | ・Nature/unknown origin<br>・Human synthesis<br>・Lab-edited<br>・Bioweapon<br>・5G<br>・E-cigarette<br>・GMO<br>・Others |
| Responsibility attribution | ・Whether a post blamed any countries/individuals for the origination of COVID-19 | ・China<br>・The United States<br>・Japan<br>・Serbia<br>・Other European countries<br>・Bill Gates<br>・Other countries (outside of the above countries)<br>・No clear responsibility attribution |
| Sources cited | ・Source is a person, thing, or place from which information comes, arises, or is quoted or referenced in a post | ・Government sources (documents, officials, organizations): Chinese government, the U.S. government, and the governments of other countries<br>・Scientific scholars<br>・Celebrities<br>・Ordinary people<br>・Foreign media<br>・Chinese media<br>・Industry/companies |

- Non-governmental organizations
- Others (outside of the above categories)
- No sources

*C. Descriptive table for each hand-coded conspiracy related variables*

The coding scheme was iteratively developed and pilot-tested with 100 randomly sampled Weibo posts from the final dataset. Six Chinese native speakers were trained and coded the posts independently. Intercoder-reliability was satisfactory, with Krippendorff's alpha for post type, origination types, responsibility attribution, and source cited at 0.834, 0.786, 0.804, 0.826, respectively. The table below provides the descriptive results of coding of each variable.

| Variable Name | Distribution | N |
| --- | --- | --- |
| **Post Types** | ・Conspiracy posts =923<br>・Debunking posts = 593<br>・Irrelevant to conspiracies or debunking =5219 | 6735 |
| **Origination Types** | ・Nature/unknown origin = 350<br>・Human synthesis =563<br>・Lab-edited = 636<br>・Bioweapon=188<br>・5G =75<br>・E-cigarette=0<br>・GMO = 7 | 1516<br>(subset to post types = conspiracy or debunking) |
| **Responsibility Attribution** | ・China = 565<br>・The United States = 712<br>・Japan =4<br>・Serbia = 1<br>・European Countries = 24<br>・Bill Gates = 2<br>・Other Countries = 36<br>・No clear responsibility attribution = 217 | 1516 |
| **Source Cited** | ・Chinese government = 47 | 1516 |

- US government = 13
- Other government = 36
- Scientists/Scholars = 438
- Celebrities = 52
- Ordinary citizens = 234
- Foreign media = 261
- Chinese media = 280
- Corporations = 2
- NGO = 5
- Others= 108
- No source is cited in a post = 268

*D. Regression table output for the baseline model described in Finding 4*

Associations of Debunking to User Participation and Mobilization - Baseline Model

|                             | Participation | Mobilization |
|-----------------------------|---------------|--------------|
|                             | (1)           | (2)          |
| Debunking posts             | −0.106*       | 0.113***     |
|                             | (0.059)       | (0.026)      |
| Male                        | −0.276***     | −0.062**     |
|                             | (0.063)       | (0.029)      |
| Influencer                  | −0.063        | 0.187***     |
|                             | (0.122)       | (0.055)      |
| Organization                | −0.674***     | 0.180***     |
|                             | (0.138)       | (0.062)      |
| Number of followers         | 0.303***      | 0.004        |
|                             | (0.019)       | (0.008)      |
| Hubei                       | −0.055        | 0.018        |
|                             | (0.140)       | (0.063)      |
| User total posts            | −0.00000      | 0.00000**    |
|                             | (0.00000)     | (0.00000)    |
| Emotion score               | −0.005*       | −0.004***    |
|                             | (0.003)       | (0.001)      |
| Emotion polarity            | −0.090        | 0.065*       |
|                             | (0.082)       | (0.037)      |
| Anger                       | 0.009         | −0.027**     |
|                             | (0.026)       | (0.012)      |
| Anxiety                     | −0.109        | 0.015        |
|                             | (0.296)       | (0.133)      |
| Sadness                     | 0.010         | −0.496**     |
|                             | (0.449)       | (0.202)      |
| Post length                 | 0.329***      | 0.090***     |
|                             | (0.043)       | (0.019)      |
| Time since posted           | 0.010***      | 0.0003       |
|                             | (0.001)       | (0.001)      |
| Constant                    | −2.623***     | −0.161       |
|                             | (0.234)       | (0.105)      |
| N                           | 1,516         | 1,516        |
| $R^2$                       | 0.283         | 0.065        |
| Adjusted $R^2$              | 0.276         | 0.056        |
| Residual Std. Error (df = 1501) | 1.084     | 0.487        |
| F Statistic (df = 14; 1501) | 42.303***     | 7.435***     |

*p < .1; **p < .05; ***p < .01

Note: we transformed our two dependent variables into the log format before we ran this model. For variable Hubei, we coded it 1 if the province field a user filled in is Hubei Province, and 0 otherwise. Conspiracy posts are the reference group for the variable "Debunking posts".

### *E. Regression table output for Figure 4*
Associations of Debunking to User Participation and Mobilization - Full Model

|                                          | Participation | Mobilization |
|------------------------------------------|---------------|--------------|
|                                          | (1)           | (2)          |
| Debunking posts                          | −0.113        | −0.014       |
|                                          | (0.273)       | (0.124)      |
| Scientists                               | −0.141        | −0.035       |
|                                          | (0.133)       | (0.061)      |
| Nonscientist sources                     | −0.105        | 0.016        |
|                                          | (0.096)       | (0.044)      |
| Nature/Unknown origin                    | −0.110        | −0.025       |
|                                          | (0.114)       | (0.052)      |
| Human-synthesis                          | 0.111         | 0.022        |
|                                          | (0.086)       | (0.039)      |
| Lab-edited                               | 0.044         | −0.020       |
|                                          | (0.087)       | (0.040)      |
| Bioweapon                                | −0.033        | −0.115**     |
|                                          | (0.113)       | (0.052)      |
| Other conspiracy types                   | −0.294        | 0.083        |
|                                          | (0.199)       | (0.091)      |
| China responsibility only                | 0.282***      | −0.018       |
|                                          | (0.089)       | (0.041)      |
| US responsibility only                   | 0.226***      | −0.024       |
|                                          | (0.086)       | (0.039)      |
| Male                                     | −0.152*       | −0.071*      |
|                                          | (0.081)       | (0.037)      |
| Influencer                               | 0.045         | 0.163***     |
|                                          | (0.122)       | (0.055)      |
| Organization                             | −0.664***     | 0.157**      |
|                                          | (0.138)       | (0.063)      |
| Number of followers                      | 0.267***      | 0.004        |
|                                          | (0.023)       | (0.010)      |
| Debunking posts:Scientists               | −0.215        | 0.226**      |
|                                          | (0.215)       | (0.098)      |
| Debunking posts:Nonscientist sources     | −0.024        | 0.040        |
|                                          | (0.187)       | (0.085)      |
| Debunking posts:Nature/Unknown origin    | −0.033        | −0.011       |
|                                          | (0.163)       | (0.074)      |
| Debunking posts:Human-synthesis          | −0.182        | −0.061       |
|                                          | (0.130)       | (0.059)      |
| Debunking posts:Lab-edited               | −0.115        | 0.015        |
|                                          | (0.137)       | (0.062)      |
| Debunking posts:Bioweapon                | −0.202        | 0.099        |
|                                          | (0.226)       | (0.103)      |
| Debunking posts:Other conspiracy types   | 0.026         | 0.060        |
|                                          | (0.282)       | (0.128)      |
| Debunking posts:Male                     | −0.301**      | 0.013        |
|                                          | (0.128)       | (0.058)      |
| Debunking posts:Number of followers      | 0.077***      | −0.001       |
|                                          | (0.027)       | (0.012)      |
| Other controls included                  |               |              |
| Observations                             | 1,516         | 1,516        |
| $R^2$                                    | 0.310         | 0.080        |
| Adjusted $R^2$                           | 0.295         | 0.060        |
| Residual Std. Error (df = 1483)          | 1.069         | 0.486        |
| F Statistic (df = 32; 1483)              | 20.846***     | 4.023***     |
| *Note:*                                  | *p<0.1; **p<0.05; ***p<0.01 ||

Note: we transformed our two dependent variables into the log format before we ran this model. For Other conspiracy types, we merged 5G and GMO into this category. For China responsibility only, we coded it 1 if a post assigned responsibility to China but not to any other entities, and 0 otherwise. For US responsibility only, we coded it 1 if a post assigned responsibility to US but not to any other entities, and 0 otherwise. For variable Scientists and Non-scientist sources, both came from a factor variable that has three levels: the reference level is no source is

cited in a post, the second level is a post cited scientists/scholars, and the third level is a post cited other non-scientist sources.

*F. Distribution of Individual Narratives*

We examined the individual narratives of each conspiracy related post by looking into the distribution of the combination of our two main hand-labelled variables: origination type and responsibility attribution since these two elements constitute the main part of a post's content. From the table below, we can see that individual narratives mainly consist of responsibility attribution to US and China when discussing the origination of COVID-19. There is not much variation in individual narratives.

| Origination Type | Individual Narratives |
|---|---|
| Nature/Unknown Origin | Nature AND China Responsible: 174<br>Nature AND US Responsible: 125<br>Nature AND China Responsible AND US Responsible: 15<br>Nature AND China AND US AND Japan Responsible: 1<br>Nature AND China AND US AND Other European Countries Responsible: 1<br>Nature AND Serbia AND US Responsible: 1<br>Nature AND US AND Other European Countries Responsible: 2<br>Nature AND Other European Countries Responsible: 5<br>Nature AND Other nonAboveCountries Responsible: 2<br>Nature AND No Clear Responsibility Attribution: 24 |
| Human-synthesis | Human AND China Responsible: 156<br>Human AND US Responsible: 253<br>Human AND China Responsible AND US Responsible: 10<br>Human AND China AND US AND Japan Responsible: 1<br>Human AND US AND Japan Responsible: 1<br>Human AND Japan Responsibility: 1<br>Human AND Other European Countries: 11<br>Human AND Bill Gates: 2<br>Human AND nonAboveCountries: 26<br>Human AND US AND nonAboveCountries Responsible: 1<br>Human AND No Clear Responsibility Attribution: 100 |
| Lab-edited | Lab AND China Responsible: 298<br>Lab AND US Responsible: 268<br>Lab AND China AND US Responsible: 13<br>Lab AND China AND US AND Japan Responsible: 1<br>Lab AND US AND Japan Responsible: 1<br>Lab AND US AND Other European Countries Responsible: 1<br>Lab AND China AND nonAboveCountries Responsible: 1<br>Lab AND No Clear Responsibility Attribution: 53 |
| Bioweapon | Bioweapon AND China Responsible: 25<br>Bioweapon AND US Responsible: 133<br>Bioweapon AND China AND US Responsible: 4 |
|  | Bioweapon AND China AND US AND Other European Countries Responsible: 1 |
|  | Bioweapon AND US AND Japan Responsible: 1 |
|  | Bioweapon AND China AND nonAboveCountries: 1 |
|  | Bioweapon AND US and nonAboveCountries: 1 |
|  | Bioweapon AND No Clear Responsibility Attribution: 22 |
| 5G | 5G AND China Responsible: 14 |
|  | 5G AND US Responsible: 13 |
|  | 5G AND Other European Countries Responsible: 4 |
|  | 5G AND No Clear Responsibility Attribution: 44 |
| GMO | GMO AND US Responsible: 6 |
|  | GMO and No Clear Responsibility Attribution: 1 |

*G. Examples of conspiracy posts and debunking posts*

The table below provides examples (in Chinese and in English translation) for conspiracy posts and debunking posts. In this paper, we define ***conspiracy posts*** as those that spread conspiracies about the origination of COVID-19. We define ***debunking posts*** to broadly include any posts that disapprove, disagree and refute such conspiracies, either with or without providing evidence. The first example of debunking post refutes conspiracy with evidence by stating that people from other countries also got COVID-19. The second and the third example under debunking posts refute conspiracy without using evidence. As we can see that the post authors denounced the conspiracy by saying it is totally nonsense. We did not fact check to verify whether the evidence used in these debunking posts are true.

We also provide the Weibo screenshots of conspiracy posts and debunking posts (in Chinese and in English) after the table.

| **Post Types** | 例子 (in Original Chinese) | **Examples** (English Translation) |
|---|---|---|
|  | 1. "*基因武器[黑线]这次新型冠状病毒很可能是美国对中国投放的。天杀美国人[怒]。*" | 1. "*Genetic weapons[blackface]. This new coronavirus is likely to be launched by the United States in China. Damn the Americans[anger].*" |
|  | 2. "*病毒真的是野味源头吗？我怎么一直感觉是台湾或者国外的故意把病菌传到中国，比如美国呢？日本呢？土耳其？澳大利亚？？？因此来限制我们的经济，限制我们的体育，限制我们的外交。是这样吗？？？！！！！有没有可能他们把人员输入进去，从而影响崛起而管制没有那么严重的武汉呢？@钟南山 @鲁健 @国家卫健委。*" | 2. "*Is the virus originated in wildlife? Why do I always feel that Taiwan or other countries deliberately spread the virus to China, such as the United States? How about Japan? Turkey? Australia? ?? So to restrict our economy, restrict our sports, restrict our diplomacy. Is that right? ? ? ! ! ! ! Is it possible that they imported infected cases into China, thus affecting the rise of Wuhan? @Zhong Nanshan @Lu Jian @Chinese National Health Commission.*" |
| **Conspiracy posts** | 3. "新冠病毒是丧失人性的实验室恶作。新冠病毒绝非野生动物携带传染人类所致，是典型的实验室人为干预、培养训化造成。万恶的敌人，企图亡我之心不死，对中国崛起耿耿于怀，丧心病狂地研究新型冠状病毒谋害华人。经过比对，敌人的阴谋昭然若揭，冠状病毒与SARS病毒形态、结构极其相似，实际上同出一宗，都是在以SARS做为病毒母体的基础上进行的干预、扩展、变异。不难发现，SARS和冠状病毒都是出自同一实验室的恶劣创作，是丧失人性的卑痞实验！中国人民必须擦亮眼睛，认清帝国主义本色，揭露敌人的阴谋诡计，与丧心病狂的敌人开展针锋相对的斗争。" | 3. *"The new coronavirus is a dehumanizing laboratory mischief, which is definitely not caused by wild animals, but a typical laboratory human intervention. The wicked enemy, trying to kill us, is notorious about the rise of China, and is frantically studying the novel coronavirus against the Chinese. After comparison, the enemy's conspiracy is rather obvious. Coronavirus and SARS virus are very similar in form and structure. In fact, they are the same. They are all interventions, expansions, and mutations based on SARS as the virus matrix. It is not difficult to find that SARS and coronavirus are both bad creations from the same laboratory, and they are dehumanizing and humble experiments! The Chinese people must keep their eyes open, recognize the true nature of imperialism, reveal the enemy's conspiracy and tricks, and wage a tit-for-tat struggle against the frenzied enemy."* |
| **Debunking posts** | 1. "又一个阴谋论破了，病毒不是针对亚洲人//////澎湃新闻：【日本政府：#钻石公主号邮轮新增41例新冠肺炎#，其中21人为日籍】日本厚生劳动省7日发布新确认感染新型冠状病毒的41人国籍，具体为21名日本人、8名美国人、5名澳大利亚人、5名加拿大人、1名阿根廷人、1名英国人。" | 1. *"**Another conspiracy theory is debunked. The virus is not aimed at Chinese**. //////The Paper: [Japanese Government: # Announced the nationalities of 41 people newly confirmed to be infected with the new coronavirus, specifically 21 Japanese, 8 Americans, 5 Australians, 5 Canadians, 1 Argentine, and 1 British]."* |
| | 2. "#武汉病毒所目前零感染# 笑死了。用脚想想，零号病人。哈哈哈哈。可真会编。" | 2. *"#Currently Zero Infection in Wuhan Institute of Virology# Laughed to death. Think with your feet, "index case"?. Hahahaha. **Totally stuff and nonsense.**"* |
| | 3. "一个做噬菌体的已毕业多年的学生，居然被阴谋论分子说是零号病人，还这么多人信……可见大家也是真的不懂……你们记住了，噬菌体是感染细菌的病毒！不能感染植物，动物！！！下次编的时候，记得找个做人源病毒的，至少是哺乳动物源吧？#武汉加油##武汉病毒所目前零感染##武汉病毒所##中国加油##造谣一张嘴,辟谣跑断腿#" | 3. *"A student majored in bacteriophage, and who graduated many years ago, was accused of being an "index case" by conspiracy theorists, and so many people believe… **It shows that everyone really doesn't understand…** You remember, bacteriophages are viruses that infect bacteria! Can not infect plants, animals!!! **When you talk nonsense next time,** remember to find a human-derived virus, at least of mammalian origin, right?"#Stay strong, Wuhan##Currently Zero Infection in Wuhan Institute of Virology#Stay strong, China##Easy to Fabricate Rumor, However, Hard to Debunk it#."* |

Conspiracy Posts Example 1:

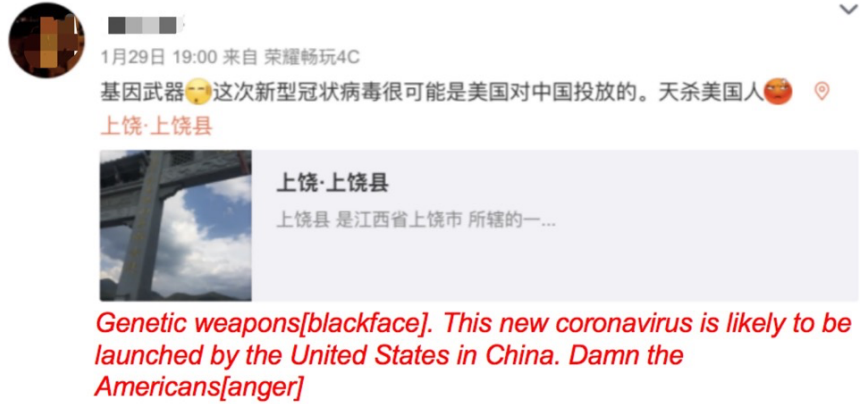

Conspiracy Posts Example 2

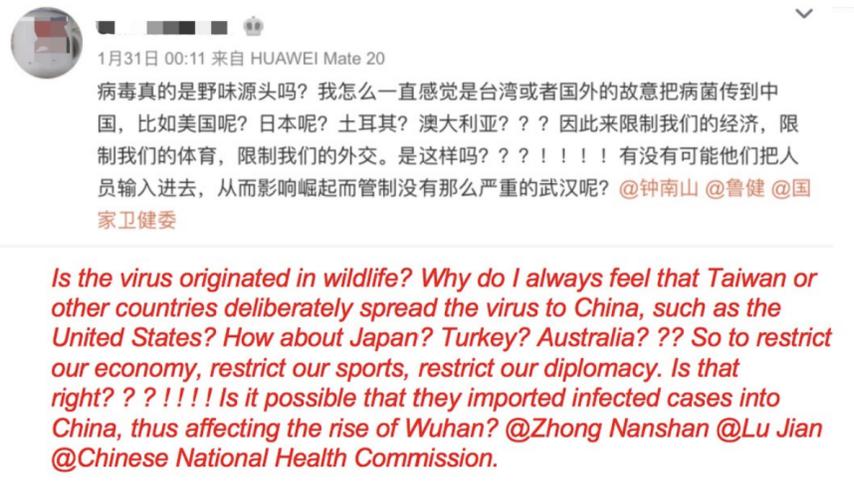

Conspiracy Posts Example 3

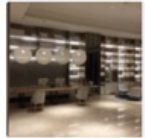

Debunking Posts Example 1

Debunking Posts Example 2

*#Currently Zero Infection in Wuhan Institute of Virology# Laughed to death. Think with your feet, "index case"?. Hahahaha. Totally stuff and nonsense.*

Debunking Posts Example 3

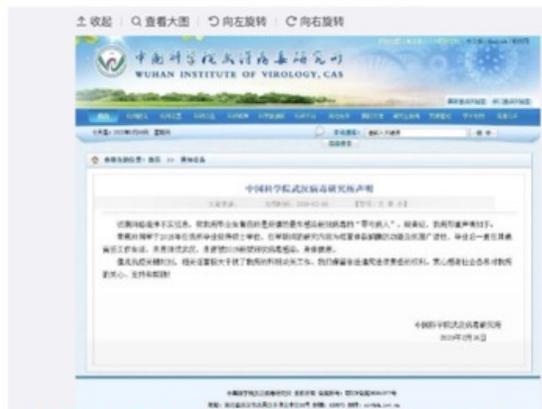

*A student majored in bacteriophage, and who graduated many years ago, was accused of being an "index case" by conspiracy theorists, and so many people believe... It shows that everyone really doesn't understand... You remember, bacteriophages are viruses that infect bacteria! Can not infect plants, animals!!! When you talk nonsense next time, remember to find a human-derived virus, at least of mammalian origin, right?"#Stay strong, Wuhan##Currently Zero Infection in Wuhan Institute of Virology#Stay strong, China##Easy to Fabricate Rumor, However, Hard to Debunk it#.*

*H. Limitation of this Study*

This study has several limitations. First, Weibo posts were collected retrospectively on May 16, 2020 and thus our dataset does not contain deleted or censored posts. However, this potential exclusion should not interfere with our conclusions as a previous study found that only 0.17% of all Weibo posts on COVID-19 were censored, and these censored posts were generally about the government's missteps in COVID-19 response, not about COVID-19 origination (Fu & Zhu, 2020). Second, our study is exploratory in nature. Findings on associations between debunking strategies and user engagement and patterns of conspiracy and responsibility attribution evolution with Sino-US conflicts should not be interpreted as causal. Finally, as our findings demonstrate, conspiracies prevalent on Chinese social media might differ significantly from those of other countries or other media systems. It will be fruitful for future research to examine major conspiracy theories emerged during COVID-19 in other countries to compare how conspiracy narratives might differ among various media systems. It will also be interesting to examine responsibility attribution by US users on Twitter. In fact, some research has shown that over 78% American blamed China for its role in spreading COVID-19 (Pew, 2020b).